# Monitoring pollution pathways in river water by predictive path modelling using untargeted GC-MS measurements

Maria Cairoli[a, 1], André van den Doel[a, 1], Berber Postma[a], Tim Offermans[a], Henk Zemmelink[b], Gerard Stroomberg[a, c], Lutgarde Buydens[a], Geert van Kollenburg[a, d, 2], Jeroen Jansen[a, 2 *]

[a] Radboud University, Department of Analytical Chemistry & Chemometrics, Institute for Molecules and Materials (IMM), Heyendaalseweg 135, 6525 AJ Nijmegen, The Netherlands.

[b] Rijkswaterstaat, Zuiderwagenplein 2, 8224 AD Lelystad, The Netherlands.

[c] RIWA Rijn, Groenendaal 6, 3439 LV, Nieuwegein, The Netherlands.

[d] Eindhoven University of Technology, Interconnected Resource-aware Intelligent Systems, Den Dolech 2, 5612AZ, Eindhoven, The Netherlands.

[1, 2] These authors contributed to this work equally.

* Corresponding author. Correspondence e-mail address: chemometrics@science.ru.nl

## Abstract

A comprehensive approach to protect river water quality is needed within the European Water Framework Directive. Non-target screening of a complete chemical fingerprint of the aquatic ecosystem is essential, to identify chemicals of emerging concern and to reveal their suspicious dynamic patterns in river water. This requires a new combination of two measurement paradigms: the path of potential pollution should be traced through the river network, while there may be many compounds that make up this chemical composition—both known and unknown. Dedicated data processing of ongoing GC-MS measurements at 9 sites along the Rhine using PARAFAC2 for non-target screening, combined with spatiotemporal modelling of these sites within the river network using path modelling (Process PLS), provided a new integrated approach to track chemicals through the Rhine catchment, and tentatively identify known and as-yet unknown potential pollutants based on non-target screening and spatiotemporal behaviour.

## Introduction

The EU Water Framework Directive (WFD) is a highly comprehensive European environmental legislation to shift the paradigm from monitoring on the level of individual target chemicals toward a holistic understanding of the aquatic ecosystem [1,2].

Current chemical monitoring takes place on the level of individual target chemicals, limiting a thorough understanding of the aquatic ecosystem [3] and the chemical diversity that affects it. While targeted analysis is invaluable in environmental health and safety monitoring, untargeted screening is essential for a holistic WFD-proof approach, as it detects as-yet unidentified chemicals, including chemicals of emerging concern [4], providing a complete chemical fingerprint of the aquatic ecosystem. Modern analytical platforms allow sensitive, untargeted detection of thousands of chemicals [5,6]. These untargeted analyses are, however, mostly used to detect and quantify priority target chemicals, and the information on molecules that are not prioritized is often left untapped [7,8].

The number of continuously released anthropogenic chemicals far exceeds what is feasible to analyze[9], therefore existing prioritization schemes mainly focus on specific contaminants [10,11], prioritized according to exposure and risk assessment [12,13]. They rarely consider spatiotemporal contamination patterns. Evaluating temporal variations in concentration patterns of yet-unidentified chemicals at a single measurement station may already provide valuable insights into which chemicals are of emerging concern [14]. However, even higher insights may be obtained by integrating spatiotemporal variation [15] among untargeted water monitoring measurements across several sampling sites.

Chemicals that enter the water system at one point could potentially be either carried through, evaporated, deposited, or broken down [16,17]: water monitoring measurements in interconnected sites along the river are necessary to characterize these behaviours [18]. Whether it exists a correlation between such chemicals depends on multiple factors, such as distance, river topology and flow connectivity [19], further than anthropogenic and natural forces [20]. Advanced statistical modelling is essential to capture such complexity: by incorporating the system knowledge into predictive modelling [21], path modelling allows unveiling causal relationships between chemicals monitored throughout the stream [20,21], ultimately revealing their spatiotemporal dynamics in the riverine.

Path modelling assumes a process consists of interrelated steps that can be modelled through a latent structure [22]. Mathematical properties of untargeted water quality data, multicollinearity and multidimensionality, make Process PLS most suited [23] to capture the water system complexity throughout connected sampling sites. As this paper will show, with the aid of suitable preprocessing and PARAFAC2 for automated feature extraction, Process PLS [21] allows for the inclusion of spatiotemporal information among different sampling sites with predictive modelling in untargeted monitoring. This makes it possible to track pollution along the watershed, monitor suspicious patterns, and even hint toward sources of contamination.

## Results and Discussion

### Data description

232-5738 samples were collected between 2012 and 2014 at regular intervals on 8 sites located on the river Rhine, and one site on the Lippe, a tributary of the Rhine. **Figure 1** specifies the sampling sites and their distances ($\Delta d$). The samples were analysed by German and Dutch water authorities with purge and trap gas chromatography-mass spectrometry (GC-MS). GC-MS is a common technique for monitoring water quality, as it detects volatiles and small organic chemicals [24–26], that may be tentatively identified by comparing their spectra to a reference database of reference spectra [27]. Calibration makes it possible to quantify target chemicals [28]; however, true identification of a measured ion involves measuring a standard for the originating chemical on the same machine, which is unfeasible for all unknowns. Process PLS aids in prioritizing unknown chemicals to be identified by adding spatiotemporal behaviour to their risk assessment. To do so, relevant chemical features need to be extracted from GC-MS samples, after only retaining samples that are synchronized over time.

### Temporal synchronization

Temporal synchronization is necessary to correlate occurrences of chemicals at different sites within the same water volume. We calculated flow durations between sites based on recorded water levels and corresponding flow times. Samples rarely exactly matched the time the volume of water reaches the next site, yet point-source contaminations are broadened downstream through diffusion so that we defined a flow time match tolerance of 1-3 hours, based on site-to-site distances. Water volumes tracked from Bad Honnef to Bimmen required matching sampling times for all in-between sites, leading to 71 water volumes sampled at all sites.

### Extracting chemical features from raw GC-MS data

We used the PARADISe software [29], based on PARAllel FActor Analysis2 (PARAFAC2), from the methods available to extract features from raw GC-MS spectra [30–32]. PARADISe divides the GC-MS spectra into retention time windows, each decomposed by PARAFAC2 into modes of mass spectra, retention time profiles and relative concentrations of PARAFAC2 components [33]. Although PARAFAC2 may handle slight retention time shifts within each window, our data required additional chromatographic alignment by Correlation Optimized Warping [34], after baseline correction with Alternating Least Squares [35].

For each time window, we selected the number of components based on fit percentage and core consistency [36]: 156 components were first selected for Bad Honnef, and 206 at the other sites. A Convolutional Neural Network built into PARADISe identifies components related to chemicals from those related to analytical artefacts [37]: after visual inspection, we discarded the components classified as baseline, finally retaining 85 components at Bad Honnef and 109 components at the other sites. The selected PARAFAC2 components are the mathematical representation of chemicals, or mixtures of chemicals, in the measured GC-MS samples. Matching the mass spectra of the PARAFAC2 components with a reference database, such as NIST, makes it possible to tentatively identify such chemicals [29].

We extracted the relative concentrations of the selected components, which we normalized to average concentrations of internal standards to eliminate differences in overall signal intensity between samples. Note that PARAFAC2 measures *relative* concentrations among samples: although comparable, they should be thought of as concentration levels, rather than absolute molar chemical concentrations, and for this reason, we will refer to them as *concentration profiles*.

**A path model of the Rhine**

Process PLS extends Partial Least Squares (PLS) regression to analyse multiple multicollinear datasets associated by a pathway [21], leading to an *outer* and *inner* model. The *outer* model assigns the extracted concentration profiles as variables to their respective sampling site ('blocks' in the data). The *inner* model (**Figure 1**) specifies relations between sampling sites, connecting each site to those directly downstream. Water enters our model in Bad Honnef and Wesel Lippe and flows to Bimmen, which is the model end-point. We treat Lobith and Bimmen as separate proximal sites ($\Delta d = 2\ km$), since on opposite sides of the river and thus might be separated in a laminar river flow. Furthermore, Orsoy Left connects to Bimmen as its first downstream site on the left side and might share the laminar flow.

A Process PLS model can be most easily interpreted using 2 statistics: $R^2$ and $P^2$, which are obtained by two consecutive steps in the modelling procedure [21]. In a Process PLS model, $R^2$ (**Figure 1**) indicates the amount of information extracted by the model from the measured variables at each site. This information is represented by sets of latent variables (LVs) that describe how much chemical variation in the water composition at a site is related to the chemical variation at other connected sites. As a result, high $R^2$ indicates high similarity in the water composition of one site with the composition of sites connected in the model specification. In the second step, the extracted chemical variation at a site is then predicted by the upstream sites that are connected to it. The power of such prediction is also quantified as explained variance, noted $P^2$, rho-squared (**Figure 1**). High $P^2$ indicates that the modelled chemical variation at a site is highly predictable from the connected upstream sites. Through $P^2$, it is possible to reveal and differentiate potential contamination patterns.

The amount of chemical variability shared among connected sampling sites, as predictive power $P^2$ and river topology, may be investigated. Low $P^2$ (**Figure 1**) indicates that chemical patterns in the predictor site are not observed at downstream sites: chemicals may have been broken down, or new chemicals may have been introduced between sites. Distances between sites are relevant in quantifying relations in their contamination: larger distances imply a higher possibility for chemicals to react, break down, or be introduced in between, hence reducing the predictive power of upstream sites. For instance, the 3 stations in Orsoy are located downstream to Bad Honnef with a distance of ~150 km (**Figure 1**): the low $P^2$ confirm that the connected sites do not share similar contamination patterns. Most of the pollution we observed had been probably introduced in between these sites.

However, the high $P^2$ between Wesel Rhine, Rees, and Lobith, which all lay on the same river bank at a shorter distance (~25 km), indicate that they share similar contamination patterns.

When a site is predicted by multiple sites, it is possible to differentiate sources of contamination, indicating where the pollution at the considered site was introduced. In our model, we can exploit this information to evaluate differences between the left and right sides of the river (in Bimmen), and the influence of a Rhine tributary on the monitored chemical variability on the main stream (in Wesel Rhine) (**Figure 1**). In Bimmen, 73.8% of all observed chemical variability in the water was included in the model ($R^2$). More than half of this variability can be predicted from information in Orsoy Left (i.e., $P^2_{BIM,ORL} = 55.0\%$) and about 43% can be predicted from Lobith ($P^2_{BIM,LOB}= 42.9\%$), which is on the right side. Although the distance between Lobith and Bimmen is much shorter (2.0 km) than between Orsoy Left and Bimmen (72.4 km), Orsoy Left predicts a higher percentage of the chemical variability in Bimmen, suggesting that a few of the monitored chemicals follow pattern on the same river bank. In Wesel Rhine, 77.4% of chemical variability was extracted from the observed data, and all of this variability was either observed in the 3 sites in Orsoy or Wesel Lippe, as the explained variances $P^2$ of those sites when predicting Wesel Rhine sum to 100%. Wesel Lippe explains 18.0%, while the 3 sites in Orsoy, which all lay on the same kilometer of the Rhine, the remaining percentage. The comparable $P^2$ ($P^2_{WSR,ORL} = 25.1\%$, $P^2_{WSR,ORM} = 28.0\%$, $P^2_{WSR,ORR} = 28.5\%$) suggest that the 3 sites in Orsoy share similar contamination patterns with Wesel Rhine. These results support the notion that the majority of the pollution observed in Wesel Rhine is coming from the main stream and introduced upstream to Orsoy, with a smaller contribution from the tributary Lippe.

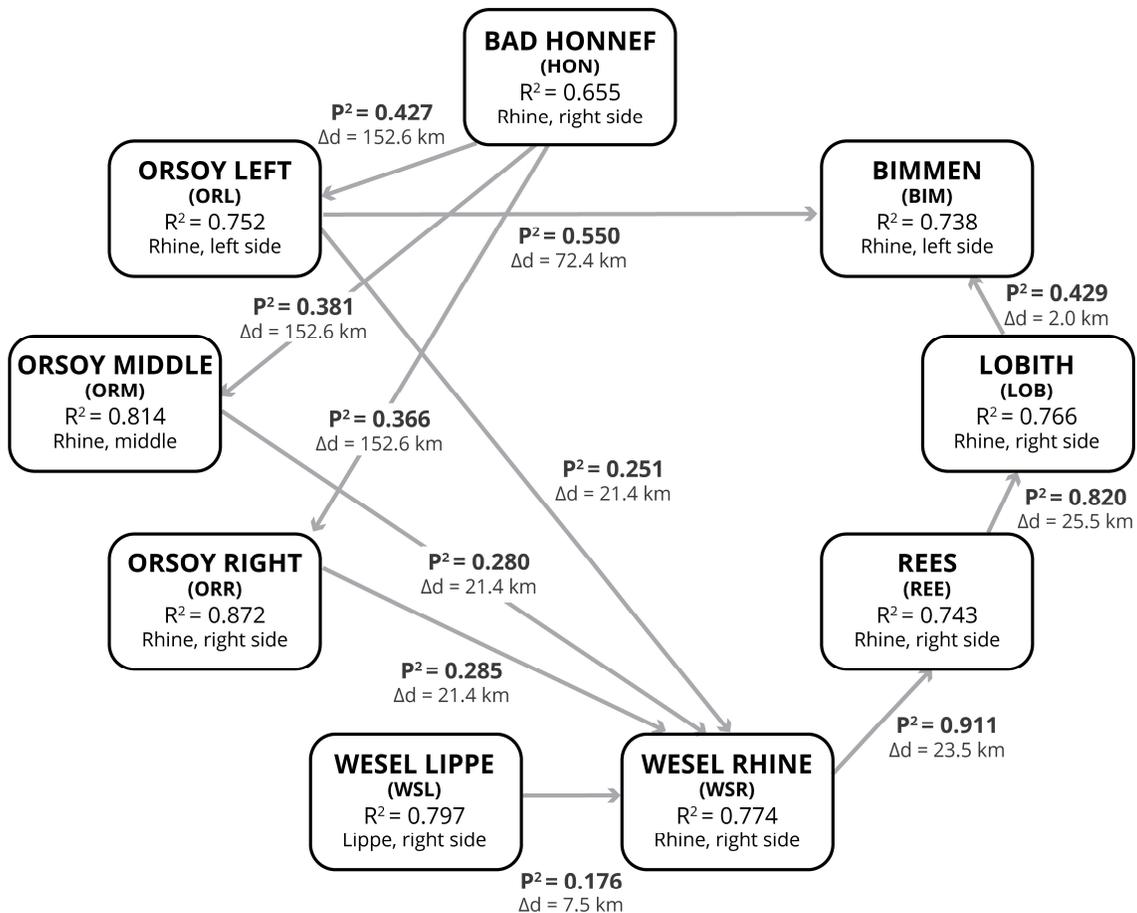

**Figure 1. Process PLS *inner* model for the 9 sampling sites on the Rhine and on the Lippe.** The arrows connect the sites that are spatially related, separated by a distance $\Delta d$. $R^2$ and $P^2$ are the explained variances in the *outer* and *inner* model.

**Tracking patterns of suspicious pollutants**

Coupling Process PLS with PARAFAC2 allows tracking **pollution patterns** of yet-unidentified chemicals, to for instance prioritize unidentified chemicals of concern, which might show suspicious behaviour. Process PLS can predict concentration profiles at a site from data collected upstream. Such predictions can be validated by comparing the concentration profiles for PARAFAC2 components as predicted by Process PLS from upstream sites with the concentration profiles measured at the to-be predicted site. The Normalized Root Mean Square Error (NRMSE), as RMSE normalized by the standard deviation of the measured profiles, is a scale-invariant quality metric for predictions at a given site. Note that quantitative comparison of Process PLS results with the measured concentration profiles requires reversion of the preprocessing steps to scale them proportionally. In this work, we were mainly interested in predicting concentration profiles from site to site, therefore we accounted for all samples in each site in our prediction.

The model predictive ability can be first evaluated by comparing measured with predicted concentration profiles for all PARAFAC2 components for each time point. We show such a comparison in Lobith and Bimmen, the model end-points on the right and left sides of the river, for a single time point (**Figure 2**). The predicted profile for Bimmen (**Figure 2b.**) shows some mismatching peaks and a higher NRMSE (0.91) than Lobith (0.15, **Figure 2a.**), indicating that not all chemicals in Bimmen were accurately predicted by the model. The model connections impact the model predictive ability for individual profiles: chemical concentration profiles in Lobith are predicted by Rees, which has no connections to other sites, whereas the prediction in Bimmen is based on Orsoy Left, optimized to predict also the chemical variability in Wesel Rhine (**Figure 1**). Such connection implies that in Bimmen not all the individual profiles were optimally predicted, and the model prediction is less reliable in this sampling site.

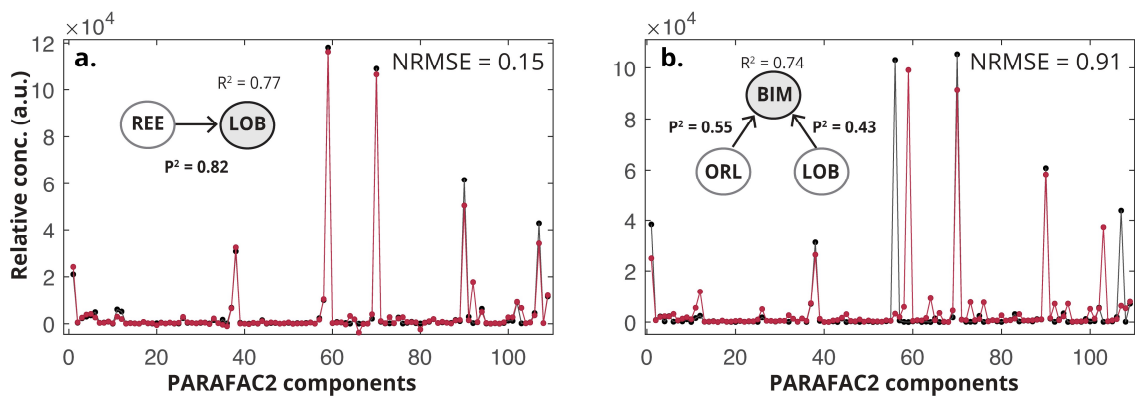

**Figure 2. Measured and predicted concentration profiles in one time point in Lobith and Bimmen.** Process PLS predicted concentration profiles (red) of the 109 PARAFAC2 components for one time point (nr. 15), and respective PARAFAC2 measured profiles (black) in Lobith (**a.**) and Bimmen (**b.**), with the fractions of the Process PLS model involving the sites.

Evaluating concentration profiles of **target chemicals** in the river connections pinpointed by $P^2$ makes it possible to validate the Process PLS results as indicative of suspicious behaviour. We show the results for cumene (benzene, (1-methylethyl)-, **Figure 3**) and MTBE (2-methoxy-2-methylpropan, **Figure 4**) as a benchmark for already monitored chemicals.

**Figure 3a.** and **3c.** display the measured concentration profiles of cumene in representative sampling sites on the right and left sides of the river. As aforementioned, the modelled variability in Bimmen is explained most strongly by Orsoy Left ($P^2_{BIM,ORL} = 55.0\%$) than by Lobith ($P^2_{BIM,LOB} = 42.9\%$). In Lobith, a high percentage of chemical variability ($P^2_{LOB,REE} = 82.0\%$) is explained by Rees. Here, $P^2$ suggests that a few chemicals follow pollution patterns on the same river bank. The measured

concentration profiles of cumene in Bimmen is much more similar to that in Orsoy Left, on the same left side, than in Lobith (**Figure 3c.**), and the concentration profile in Lobith to that in Rees (**Figure 3a.**), on the same right side. Hence, $P^2$ captures such variability. **Figure 3b.** and **3d.** show the predicted concentration profiles of cumene in Lobith and Bimmen respectively. Although not accurate due to some mismatching peaks, in Lobith the predicted profile has a lower NRMSE (NRMSE = 0.81) than Bimmen (NRMSE = 0.99), confirming the lower model predictive ability in Bimmen discussed above (**Figure 2**). We can finally verify that the PARAFAC mass spectrum of cumene corresponds to its NIST reference (**Figure 3e.**), although the matching probability is not particularly high.

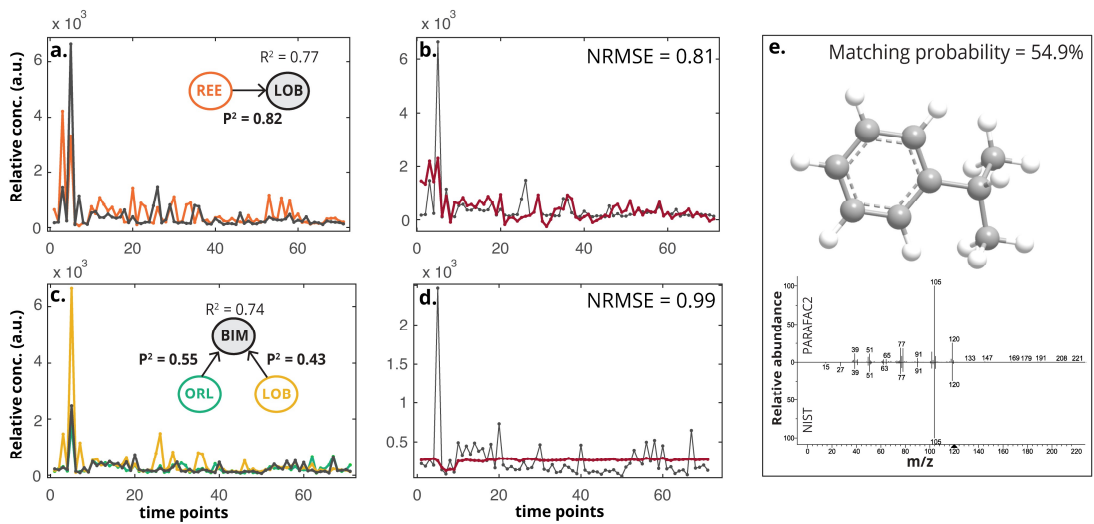

**Figure 3. Measured and predicted concentration profiles of cumene on the right and left sides of the river. a.** Cumene concentration profile measured by PARAFAC2 in Rees (orange) and Lobith (black). **c.** Measured cumene concentration profile in Orsoy Left (green), Lobith (yellow), and Bimmen (black). **b.; d.** Measured (black) and predicted (red) cumene concentration profile in Lobith (**b.**) and Bimmen (**d.**). **e.** Matching of PARAFAC2 cumene mass spectrum with NIST reference (matching probability = 54.9%).

We considered MTBE as target chemical to evaluate the influence of multiple pollution sources in Wesel Rhine. **Figure 4a.** displays the measured concentration profiles of MTBE in Orsoy Left, Orsoy Middle, Orsoy Right, Wesel Lippe and Wesel Rhine. The MTBE profile in Wesel Rhine overlaps with the profiles measured in the 3 stations in Orsoy, which $P^2$ in their prediction to Wesel Rhine is comparable ($P^2_{WSR,ORL} = 25.1\%$, $P^2_{WSR,ORM} = 28.0\%$, $P^2_{WSR,ORR} = 28.5\%$) and strongly differs from Wesel Lippe, which $P^2$ is lower ($P^2_{WSR,WSL} = 18.0\%$). We can thus conclude that MTBE in Wesel Rhine was introduced from Orsoy, and not from Wesel Lippe. To substantiate the possibility to employ the model in Early Warning Systems for preventive protection of river water quality, we can now predict whether such contaminant continues travelling downstream to Rees. **Figure 4b.** shows that the model accurately predicts the concentration profile of MTBE in Rees with NRMSE = 0.67. The matching of the PARAFAC2 mass spectrum with the NIST database confirms that the monitored chemical is MTBE (**Figure 4c.**). Such results confirm that, by differentiating chemical sources through descriptive statistics, Process PLS enables prioritization of sources of contamination and identification of suspicious patterns of pollution, further allowing to predict contamination downstream.

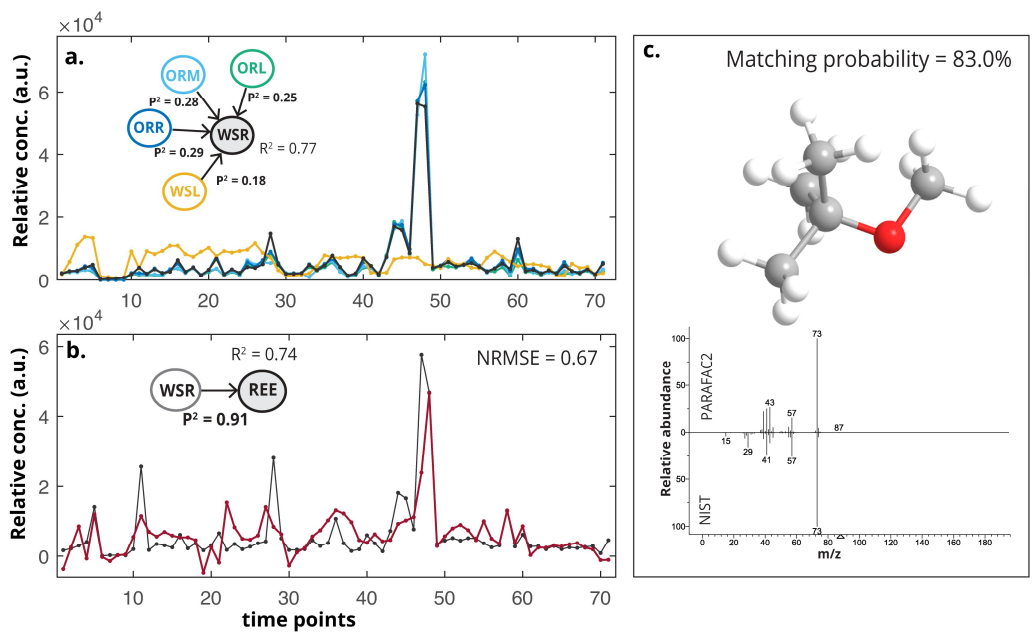

**Figure 4. Measured concentration profiles of MTBE in the Rhine and the Lippe, and profile prediction downstream. a.** MTBE concentration profile measured by PARAFAC2 in Orsoy Left (green), Orsoy Middle (light blue), Orsoy Right (blue), Wesel Lippe (yellow) and Wesel Rhine (black). **b.** Measured (black) MTBE concentration profile in Rees and predicted profile by Wesel Rhine (red). **c.** Matching of PARAFAC2 MTBE mass spectrum with NIST reference (matching probability = 83.0%).

Further analyses enable us to prioritize and tentatively annotate **non-target chemicals** in the GC-MS data. **Figure 5a.** shows the measured concentration profile of a chemical that has comparable concentration profiles in the 3 stations in Orsoy and in Wesel Rhine, but not in Wesel Lippe. As MTBE, this chemical was introduced upstream to Orsoy and not from the Lippe, substantiating what indicated by $P^2$. Unlike MTBE, we observe higher concentration peaks in Wesel Rhine, indicating the occurrence of a pollution event between Orsoy and Wesel Rhine. By matching its PARAFAC2 mass spectrum with the NIST database (**Figure 5a.**) we could tentatively identify such chemical as cyclopentadiene, a pollutant that can be found in environmental samples [26]. The prediction in the downstream station Rees is not accurate (**Figure 5b.**, NRMSE = 0.96). The measured values show a peak 4 times higher than the peak in Wesel Rhine, which is a clear indication of a pollution event that happened in between the two sampling sites. By design, the used model is not able to predict such intermediate pollution, but as shown here, the model can be used to tentatively identify the source of pollution (in between the sites) and provides evidence that this unknown chemical should be further investigated.

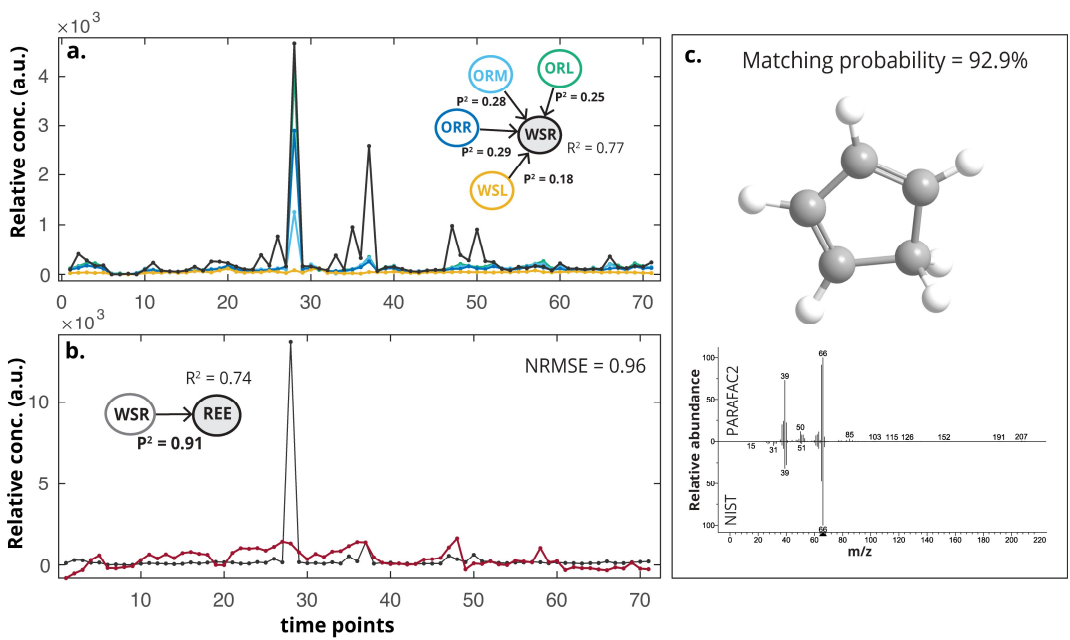

**Figure 5. Measured concentration profile of untargeted chemical in the Rhine and in the Lippe, Process PLS prediction downstream and tentative identification. a.** Untargeted chemical concentration profile measured by PARAFAC2 in Orsoy Left (green), Orsoy Middle (light blue), Orsoy Right (blue), Wesel Lippe (yellow) and Wesel Rhine (black). **b.** Measured (black) chemical concentration profile in Rees and predicted profile by Wesel Rhine (red). **c.** Matching of PARAFAC2 mass spectrum of untargeted chemical with NIST reference spectrum of 1,3-cyclopentadiene (matching probability = 92.9%).

We finally investigated chemicals which patterns repeat consistently throughout the sampling sites, looking at the connections involving Wesel Rhine, Rees and Lobith, where high $P^2$ (**Figure 1**) indicate a high similarity in the pollution patterns.

**Figure 6a.** shows the measured concentration profiles in Wesel Rhine and Rees of a chemical that we tentatively annotated as acetonitrile with 93.5% matching NIST probability (**Figure 6c.**). Acetonitrile is a commonly used solvent and organic synthesis intermediate, and exposure at sufficiently high doses can cause cyanide poisoning; our model shows that its pattern is consistently repeating throughout the sampling sites. Recent studies [14,38] demonstrated that anthropogenic contaminants show a repeating pattern of peaks through time, which substantiates implementing monitoring of acetonitrile in the river. We successfully predicted the concentration profile of acetonitrile in Lobith from Rees (**Figure 6b.**).

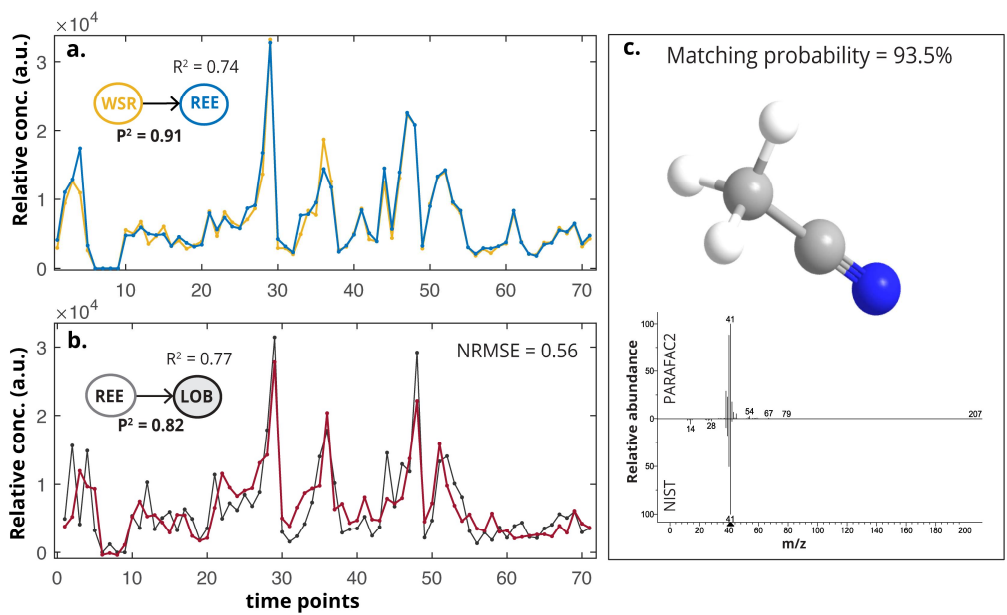

**Figure 6. Measured concentration profile of untargeted chemical on the right side of the Rhine, Process PLS prediction downstream and tentative identification. a.** Chemical concentration profile measured by PARAFAC2 in Wesel Rhine (yellow), and Rees. **b.** Measured (black) chemical concentration profile in Lobith and predicted profile by Rees (red). **c.** Matching of PARAFAC2 untargeted chemical mass spectrum with NIST reference spectrum of acetonitrile (matching probability = 93.5%).

**Table 1** reports an overview of the chemicals that we were able to tentatively identify from Orsoy Left to Bimmen, for which we were provided with a table of already monitored chemicals. The PARAFAC2 analysis extracted some (9) chemicals as multiple components, albeit with varying matching probabilities. If these are indeed the same chemicals, this may lead to multicollinearity in the data. While Process PLS is designed to analyze multicollinear data, certain model details (*e.g.*, variable loadings) should be interpreted carefully, but this is outside of the scope of the current paper.

**Table 1. Tentatively identified chemicals.** Chemicals tentatively identified by comparison of PARAFAC2 mass spectra with NIST reference database. The table reports the percentage of matching probability, as well as the information on whether the chemical was already monitored in the river.

| PRAFAC2 COMPONENT | NIST matching | Matching probability (%) | Already monitored? |
|---|---|---|---|
| 1 | Carbon dioxide | 73.5 | No |
| 6, 10 | Propane, 2-fluoro | 80.7, 54.2 | No |
| 11 | Acetonitrile | 93.5 | No |
| 12 | Acetic anhydride | 44.1 | No |
| 16 | Cyclopropane, 1, 1-dimethyl- | 63.0 | No |
| 17 | Dimetyl sulfide | 93.2 | No |
| 19 | 1, 3- cyclopentadiene | 92.9 | No |
| 20, 38, 41, 45, 48 | Methylene chloride | 98.9, 97.5, 97.5, 97.5, 97.2 | No |
| 21 | Carbon disulfide | 70.8 | No |
| 22 | Isoprene | 96.3 | No |
| 23 | Ethanol, 2-(trimethylsilyl)- | 51.7 | No |
| 26, 32 | Propane, 2-methoxy-2-methyl (MTBE) | 83.0, 89.3 | Yes |
| 29 | 3-butyn-2-ol | 65.9 | No |
| 34, 40 | Ethylene, 1,2-dichloro-(E) Ethylene, 1,2-dichloro-(Z) | 66.7, 68.5 18.5, 11.0 | Yes |
|  | Ethylene, 1,1-dichloro- | 14.2, 20.1 |  |
| 35 | Ethyl acetate | 68.1 | No |
| 37, 43 | Trichloro methane | 98.0, 80.5 | Yes |
| 42 | Cyclopentane, methyl- | 25.8 | Yes |
| 44, 51 | Benzene | 81.8, 77.6 | Yes |
| 47 | Cyclohexene | 59.2 | Yes |
| 49 | Trichloroethylene | 98.6 | Yes |
| 57, 67 | Toluene | 23.5, 27.7 | Yes |
| 58 | 3-Furanmethanol | 58.8 | No |
|  | 2-Furanmethanol | 26.8 |  |
| 62 | Tetrachloroethylene | 98.8 | Yes |
| 72, 74 | Styrene | 38.7, 42.0 | Yes |
| 75 | o-Xylene | 42.2 | Yes |
|  | Benzene, 1,3-dimethyl | 23.0 |  |
|  | p-Xylene | 25.3 |  |
| 77 | Benzene, (1-methylethyl)- | 54.9 | Yes |
| 81, 86 | α-Methylstyrene | 60.0, 58.6 | No |
| 89 | Benzene, 1,2,3-trimethyl- | 22.7 | Yes |
|  | Benzene, 1,2,4-trimethyl- | 20.9 |  |
|  | Mesitylene | 15.6 |  |
| 96 | Benzene, 1,2,4,5-tetramethyl- | 19.4 | No |
|  | Benzene, 1,2,3,5-tetramethyl- | 16.4 |  |
|  | Benzene, 1,2,3,4-tetramethyl- | 14.5 |  |

Overall, we tentatively identified 16 chemicals that were not previously monitored in the Rhine. Several chemicals could not be identified, and thus still remain unknown. **Figure 7a.-7c.** show the measured concentration profile, the predicted profile, and the mass spectrum of a still-unknown chemical with a consistent pattern throughout the river, which thus might represent a suitable candidate for prioritization.

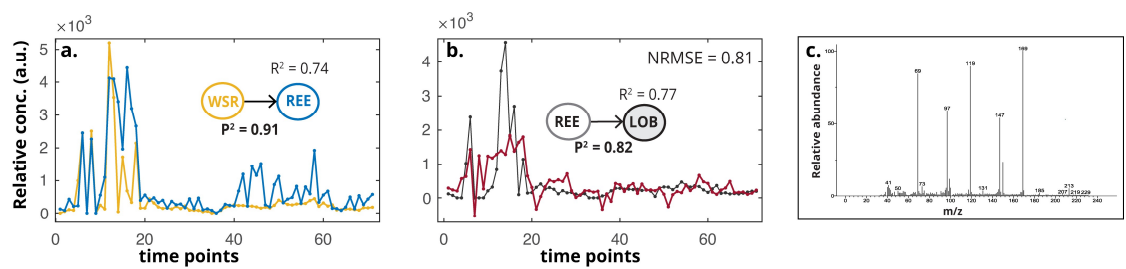

**Figure 7. Measured concentration profile of unidentified untargeted chemical on the right side of the Rhine, and profile prediction downstream. a.** Measured untargeted chemical concentration profile in Wesel Rhine (yellow), and Rees (blue). **b.** Measured (black) chemical concentration profile in Lobith and predicted profile by Rees (red). **c.** PARAFAC2 mass spectrum of unidentified untargeted chemical.

### Opportunities and challenges of predictive monitoring on untargeted analysis

Process PLS combined with PARAFAC2, temporal and spectral alignment, provides a breakthrough way to analyse pollution throughout the Rhine watershed, enabling detection and prediction of concentration profiles of unidentified chemicals between monitoring sites that remain elusive from conventional analyses on single chemicals and/or measurement sites. While not perfect for all chemicals, the approach enabled us to find, explain, and predict suspicious spatiotemporal patterns of known and unknown chemicals at different measurement sites, tentatively identifying chemicals of emerging concern.

Several chemicals cannot be assessed within the current prioritization schemes due to a lack of sufficient information; gathering this information might require considerable effort [39]. By inspecting suspicious spatiotemporal patterns throughout multiple sites and differentiating pollution sources, our approach adds a complementary priority attribute of great environmental relevance to select (unidentified) chemicals to be investigated, to finally take effective mitigation measures.

To obtain robust results, we chose to consider only repeated measurements of the same water volume travelling downstream in all sampling sites. This represented a limiting factor in our analysis due to a lack of harmonization between sampling schedules at different monitoring sites—for which there has until now been no imminent need. Harmonizing sampling times according to river flow will greatly increase the amount of data available to monitoring approaches like ours. The strength of suspicious spatiotemporal patterns may be discovered in as-yet non-priority compounds. The possibilities to analyze disharmonious measurements in many sampling sites could also be investigated in future work.

Path modelling supports an integrated modelling approach in which different parts of the ecosystem are combined in a single model [40–42], which further substantiates the WFD holistic approach. The Process PLS Latent Variable representation allows for fusing data from complementary sources, including other chemical platforms (such as LC-MS), meteorological conditions, and point and diffuse discharges. This will allow a wider range of chemical, metabolic and ecological patterns in the river water to be studied.

Ensuring sustainable clean water on a global scale is the major ambition set by the Sustainable Development Goal (SDG) 6 [43]. Early Warning Systems are nowadays required in drinking water management to predict the impact of contamination in real-time, avoiding further pollution and regularly protecting river water quality [44,45]. GC-MS instruments can measure water chemistry in near real-time [46], and path modelling can investigate and predict its variations among several, interconnected, sampling sites. Further extending this integrated approach by including online measurements will support the model implementation as an automated on-line sensor in river monitoring to detect sudden changes in river water quality downstream, upholding the SDG 6.

## Concluding remarks

We proposed path modelling with Process PLS as a breakthrough method to combine untargeted water quality data with spatiotemporal information for chemical prioritization in river water quality analysis. We were able to differentiate pollution sources and confirm the suspicious behaviour of known pollutants, giving insights into other chemicals with similarly suspicious behaviour, including those chemicals that were yet unidentified, such as cyclopentadiene and acetonitrile, and others that remain unknown. Due to its intrinsic predictive ability, path modelling offers the opportunity to develop evidence-based early warnings for downstream pollution events in comprehensive watershed management.

## Methods

### Dataset

Water samples were collected at 8 sampling sites along the river Rhine and at one site on the river Lippe, which is a tributary of the Rhine. **Table 2** reports the number of samples collected for each site, together with the river kilometer. Purge and trap gas chromatography-mass spectrometry (GC-MS) measurements of such samples were performed in full-scan mode by German and Dutch authorities: samples for Bad Honnef were measured in the monitoring station of Bad Honnef, while samples from the remaining 8 sampling sites were measured at the International Measuring station Bimmen-Lobith (IMBL).

**Table 2. Sampling sites description.**

| Sampling site | Sampling site name | Number of samples | Location on the stream (km) | Flow tolerance time (between sampling sites) |
|---|---|---|---|---|
| 1 | Bad Honnef (HON) | 5515 | 640.0 (Rhine, right side) | 3 h (1 − 3) |
| 2 | Orsoy Left (ORL) | 254 | 792.6 (Rhine, left side) | 1 h (2 - 3) |
| 3 | Orsoy Middle (ORM) | 260 | 792.6 (Rhine, middle) | 1 h (3 - 4) |
| 4 | Orsoy Right (ORR) | 258 | 792.6 (Rhine, right side) | 3 h (4 − 5) |
| 5 | Wesel Lippe (WSL) | 307 | 3.56 (Lippe, right side) | 1 h (5 − 6) |
| 6 | Wesel Rhine (WSR) | 232 | 814.0 (Rhine, right side) | 3 h (6 − 7) |
| 7 | Rees (REE) | 238 | 837.5 (Rhine, right side) | 3 h (7 − 8) |
| 8 | Lobith (LOB) | 3481 | 863.0 (Rhine, right side) | 2 h (8 − 9) |
| 9 | Bimmen (BIM) | 5738 | 865.0 (Rhine, left side) | - |

### Preprocessing: temporal synchronization and chromatographic alignment

*Temporal synchronization*

Temporal synchronization is necessary to correctly correlate chemicals monitored at multiple sampling sites over time, by tracking the same column of water in each site. We estimated the time the river water flows from one site to another according to a table of recorded water levels (cm) and flow time values (h), provided by the German Federal Institute of Hydrology through personal correspondence. Flow time values were not available for all the water samples, thus we fitted a third-degree polynomial to the available flow times and water levels in each corresponding river section and we used the obtained fitting coefficients, together with the available water levels, to estimate the flow times for all the samples. We estimated the time at which the water travelled throughout connected sampling sites according to **equation (1)**, finally retaining only the matching samples.

$$t_B = t_A + t_f \tag{1}$$

where $t_A$ is the time the water sample was collected at station A, $t_B$ the corresponding time at the connected station B and $t_f$ is the flow time value resulting from the previous extrapolation, which accounts for an additional tolerance time of 1-3 hours, estimated according to the distance between the sampling sites, and reported in **Table 2**. We repeated this operation consecutively from Bad Honnef (site 1) to Bimmen (site 9), retaining any time the matching samples. In total, we selected 71 water samples synchronized among all the sampling sites. We employed such samples in the next analysis step, which consisted in extracting relevant chemical features from the GC-MS spectra, after spectral preprocessing.

*Chromatographic alignment*

GC-MS data are often affected by spectral artefacts that hinder proper feature extraction, such as background noise, overlapping and shifting peaks, which might derive from experimental conditions, variations in the chromatogram or the mass detector [47]. Although PARAFAC2 may handle such artefacts to a certain extent [33], our data required three preliminary spectral preprocessing steps.

We aligned the GC-MS spectra according to the Total Ion Current (TIC). To compensate for different time-scales of TIC in retention times, the first step required aligning the TIC values to an equally spaced vector $\mathbf{v}_{ref}$ of $r$ retention times, defined as (**equation (2)**):

$$\mathbf{v}_{ref} = [\ t_0 = 0\ ,\ ..\ ,t_r = \max(rt)]$$ (2)

where $\max(rt)$ is the maximum retention time value assessed among the samples. For each sample, we assigned each retention time to its closest value in the reference vector. Due to the higher resolution of the spectra measured in Bad Honnef compared to the other measurements, we defined a reference vector of 7000 time points for Bad Honnef, and a vector of 4300 time points for the remaining 8 sampling sites. Due to this difference, we analyzed the chromatograms in Bad Honnef separately from the ones collected in the remaining sampling sites. The second and third preprocessing steps consisted of baseline correction through Alternating Least Squares [35], followed by Correlation Optimized Warping (COW)[34] to correct for peak shifting.

**PARAFAC2 to extract relevant chemical features**

We employed the software PARADISe [29], based on PARAllel FActor Analysis2 (PARAFAC2), to extract pure mass spectra, elution profiles and concentration profiles from the GC-MS spectra. PARAFAC2 allows to deconvolute pure mass spectra of peaks and to integrate areas of deconvoluted peaks (extracting relative concentrations) for all samples simultaneously while handling co-eluted, retention time shifted and low signal-to-noise ratio chromatographic peaks [48,49].

For each sample $k$, PARAFAC2 decomposes the GC-MS matrix $\mathbf{T}_k$ ($I$ x $J$), with $I$ mass spectra and $J$ retention times, in three loading matrices, each corresponding to 'modes' of the GC-MS spectra [33] (**equation (3)**):

$$\mathbf{T}_k = \mathbf{AD}_k(\mathbf{B}_k)^T + \boldsymbol{E}_k$$ (3)

Where $\mathbf{A}$ ($I$ x $F$) is the matrix of the mass spectra-mode of the resolved analytes $F$, which are the PARAFAC2 components. $\mathbf{D}_k$ ($F$ x $F$) is the diagonal matrix that holds the $k^{th}$ row of the sample-mode loadings matrix $\boldsymbol{C}$, which holds the chemical relative concentration profiles. $\mathbf{B}_k$ ($I$ x $F$) is the matrix of the elution profiles-mode for each component $F$, and the matrix $\boldsymbol{E}_k$ holds the model residuals [33].

PARADISe divides the chromatogram into retention time windows, and builds a PARAFAC2 model for each investigated region, after defining the number of PARAFAC2 components for each model. We manually selected the retention time intervals, and we imposed the non-negativity constraint on all the models, which implies positive mass-loadings. We constructed the models from 1 to 7 components as software-default, and for each model we selected the number of optimal components according to the criteria of core consistency and fit percentage [36]. Due to the difference in resolution, we processed the spectra measured in Bad Honnef separately from the other sampling sites in PARADISe. This led to select a total of 185 PARAFAC2 components for Bad Honnef and 206 for the remaining sampling sites. Once the models are determined, a Convolutional Neural Network built into the software classifies the components as baseline or (mixture of) chemicals [37]. We excluded the components classified as baseline to only select components corresponding to chemicals [29]. After such selection, we finally obtained 85 components for Bad Honnef and 109 for the remaining sampling sites. The resolved peaks of such components could be then tentatively identified using their deconvoluted mass spectra and the NIST reference database.

We extracted relative concentration profiles for each component in each sampling site, obtaining 9 different concentration matrices, of dimension (71 x 85) for Bad Honnef, and (71 x 109) for the 8 remaining sampling sites. Before Process PLS, we standardized such profiles to average concentrations of internal standards (deuterochloroform, toluene-d8, chlorobenzene-d5, dichlorobenzene-d4, naphthalene-d8) to eliminate differences in overall signal intensity between samples, a standard practice in analyzing GC-MS data.

**Process PLS to track pollution patterns**

Process PLS is a path modelling tool that makes it possible to find relationships between multivariate data matrices connected throughout a structure. The workhorse of Process PLS is the Partial Least Squares method (SIMPLS algorithm [50]), which is performed in two rounds to find the relationships between the measured data matrices. A Process PLS model consists of 2 sub-models: the *outer* and *inner* models. In the *outer* model, the relationships between the measured variables and the Latent Variables estimated by PLS are explored, while the *inner* model is used to analyze the relationships between the blocks of the estimated Latent Variables [21]. A brief description of the method is here provided; we refer to the reference article [21] for further details.

The first step in Process PLS consists in specifying the model: the *outer* model is specified by assigning the measured variables to the corresponding blocks in the model. The *inner* model is then specified, by connecting the blocks according to defined relationships [21]. We specified the *outer* model by assigning the concentration profiles extracted by PARAFAC2 to the sampling sites (our model 'blocks'), and we specified the *inner* model by connecting the sampling sites according to the river topology. After specifying the model, the *outer* and *inner* model are computed, and the defined relationships can be interpreted through 2 statistics: the explained variances $R^2$ and $P^2$.

*Outer model*

In the *outer* model, the first round of PLS is performed: the variables measured in each block predict the variables of the blocks connected through the inner model, obtaining a set of Latent Variables $\hat{\xi}_m$ for each block $m$. The number of estimated Latent Variables varies per block, and is computed by double cross-validation. The amount of significant information extracted by the Latent Variables is quantified, for each block $m$, by the explained variance $R^2$, estimated as (**equation (4)**):

$$R_m^2 = \frac{trace(\mathbf{L}_m^T \mathbf{L}_m)}{N-1} \tag{4}$$

for all the blocks that predict other blocks (all the sites except Bimmen in our model), and as (**equation (5)**):

$$R_m^2 = \frac{trace(\mathbf{Q}_m^T \mathbf{Q}_m)}{N-1} \tag{5}$$

for the block that only functions as a target (Bimmen in our model). In **equation (4) and (5)**, $\mathbf{L}_m$ is the **X**-loadings matrix for block $m$, $N$ is the number of observations, and $\mathbf{Q}_m$ is the **Y**-loadings matrix for block $m$ [21]. **X** is the matrix of predictor variables, while **Y** is the matrix or variables that are being predicted.

*Inner model*

After each set of Latent Variables $\hat{\xi}_m$ per block $m$ is estimated, the second round of PLS is performed, where each block $m$ is predicted by the $n$ blocks that are connected to it in the *inner* model. The explained variance of such prediction is calculated by subtracting the sum of squared errors between the estimated LVs $\hat{\xi}_m$ and the PLS prediction $\chi_m B_m$ from the total sum of squares of $\hat{\xi}_m$ [21] (**equation (6)**).

$$P_m^2 = 1 - \sum_{i=1}^{N}(\hat{\xi}_{mi} - \chi_{mi}B_{mi})^2 = 1 - SS(\hat{\xi}_m - \chi_m B_m) \tag{6}$$

Where $SS$ indicates the sum of squares, $\chi_m$ is the regression matrix, which combines the Latent Variables of the $n$ predictor blocks, according to the connections defined in the *inner* model [21] (**equation (7)**):

$$\chi_m = [\hat{\xi}_{m,1}, \dots, \hat{\xi}_{m,n}] \tag{7}$$

And $B_m$ is the matrix of the PLS regression coefficients (**equation (8)**):

$$B_m = [B_{m,1}, \dots, B_{m,n}] \tag{8}$$

Since the variance obtained in **equation (6)** is not specific for single connections between blocks, the partial explained variance for each specific predictor block $z$ is finally computed as (**equation (9)**):

$$P_{m,z}^2 = P_m^2 * \frac{SS(\hat{\xi}_m - \chi_m B_m)}{\sum_{q=1}^{n} SS(\hat{\xi}_m - \chi_{m,q}B_{m,q})} \tag{9}$$

which is indicated as $P^2$ ('rho-squared') in the connections in **Figure 1**.

After the *outer* and *inner* model are computed, individual concentration profiles can be predicted from the **X**- and **Y**- loadings and scores PLS matrices, accounting for the steps of autoscaling, mean centering and block scaling performed by the Process PLS algorithm [21].

**Software**

Matlab R2020b, PARADISe (version 3.9), Python 3.9.7.

### Author contribution

**Maria Cairoli**: Validation, Formal analysis, Investigation, Data curation, Writing - original draft, Visualization. **André van den Doel**: Conceptualization, Methodology, Validation, Investigation, Writing - original draft, Supervision, Visualization. **Berber Postma**: Methodology, Validation, Formal analysis, Investigation, Data curation. **Tim Offermans**: Formal analysis, Investigation, Data curation, Writing - review & editing. **Henk Zemmelink**: Resources, Project administration, Funding acquisition. **Gerard Stroomberg**: Supervision, Resources, Project administration, Funding acquisition. **Lutgarde Buydens**: Supervision, Project administration, Funding acquisition. **Geert van Kollenburg**: Conceptualization, Methodology, Software, Validation, Writing - review & editing, Visualization, Supervision. **Jeroen Jansen**: Methodology, Writing - review & editing, Supervision, Project administration, Funding acquisition.

### Competing interests statement

The authors declare that they have no known competing financial interests or personal relationships that could have appeared to declare influence the work reported in this paper.

## Acknowledgements

This project is co-funded by TKI-E&I with the supplementary grant 'TKI-Toeslag' for Topconsortia for Knowledge and Innovation (TKI's) of the Ministry of Economic Affairs and Climate Policy. The authors thank all partners within the project 'Measurement for Management (M4M)' (https://ispt.eu/projects/m4m/), managed by the Institute for Sustainable Process Technology (ISPT) in Amersfoort, the Netherlands. Part of this work was supported by NWO-TA-COAST (grant 053.21.114). Part of this work was supported by the ECSEL Joint Undertaking (JU) under grant agreement No 826589.